\documentstyle[aps,twocolumn]{revtex}
\def\k{\kappa}
\def\r{\rho}
\def\s{\sigma}
\def\t{\tau}
\def\e{\epsilon}
\def\b{\beta}
\def\del{\delta}
\def\D{{D}}
\def\n{{\bf n}}
\def\m{{\bf m}}
\def\eth{\hbox{$\partial$\kern-0.45em\raise0.55ex\hbox{--}}}
\def\thorn{I\kern-0.4em\raise0.35ex\hbox{\it o}}
\def\Ph#1#2{\Phi_{#1#2}}
\def\th{\theta}

\def\d{{\rm d}}
\def\p{\partial}
\def\w{\wedge}

\def\bthi{{\theta}^i}
\def\bthj{{\theta}^j}
\def\bei{{\bf e}_i}
\def\bej{{\bf e}_j}
\def\be3{{\bf e}_3}
\def\bfi{{\bf f}_i}

\def\g{{\bf g}}
\def\gd{{\bf g}}

\def\h{{\bf h}}
\def\thAB{\theta^A{}_B}
\def\thAC{\theta^A{}_C}
\def\thCB{\theta^C{}_B}
\def\ric{{\rm Ric}}
\def\S{\Sigma}
\def\kg{\kappa_g}
\newtheorem{prop}{Proposition}[section]

\newtheorem{lem}[prop]{Lemma}

\begin{document}
\twocolumn[\hsize\textwidth\columnwidth\hsize\csname @twocolumnfalse\endcsname

\title{How much negative energy does a wormhole need?}
\author{Daisuke Ida\dag and Sean A. Hayward\ddag}
\address{
\dag Department of Physics, Kyoto University, Kitashirakawa,
Sakyo-ku, Kyoto 606-8502, Japan\\ 
{\rm E-mail address: ida@tap.scphys.kyoto-u.ac.jp}\\
\ddag Yukawa Institute for Theoretical Physics, Kyoto University,
Kitashirakawa, Sakyo-ku, Kyoto 606-8502, Japan\\
{\rm E-mail address: hayward@yukawa.kyoto-u.ac.jp}}
\date{\today}
\maketitle
\begin{abstract}
It is known that traversible wormholes require negative energy density.
We here argue how much negative energy is needed for wormholes,
using a local analysis which does not assume any symmetry.
and in particular allows dynamic (non-stationary) but non-degenerate wormholes.
We find that wormholes require two constraints on the energy density,
given by two independent components of the Einstein equation.
keywards: traversible wormhole, energy condition
\end{abstract}
]
\section{Introduction}
A spacetime wormhole is usually introduced as a topological handle
connecting two universes or distant places in
a universe. 
The research of wormholes became active in the last decade
since the work of Morris and Thorne \cite{MT}.
A remarkable result is that, assuming Einstein gravity, 
the weak energy condition is violated
at the wormhole throat.
Though their analysis was restricted to the static, 
spherically symmetric case,
other examples suggest that the violation of energy conditions, 
particularly the null energy condition,
is a general property of traversible wormholes \cite{shell}--\cite{dwh}.
Since any realistic gravitational source is believed to satisfy the
null energy condition at the classical level,
one might look to effects of quantum field theory 
to maintain a wormhole.
In this context, Hochberg et al. \cite{semiclass}
have presented a self-consistent
wormhole solution of semiclassical Einstein gravity.
Another way is to consider alternative gravitational theories, such as 
higher-derivative theories \cite{R2,higher} 
or the Brans-Dicke theory \cite{BD}.

Recently, general frameworks for dynamic wormholes
have been suggested.
While in static spacetimes, a wormhole may be defined by 
a closed stable minimal surface embedded in the hypersurface
orthogonal to the timelike Killing vector field,
Hochberg and Visser \cite{necdwh,dwhats} have generalized this
concept for general spacetimes and defined a wormhole
by a stable minimal surface on a null (lightlike) hypersurface.
They have also shown that the null energy condition
is violated at the wormhole throat in general.
On the other hand, one of the present authors has defined a dynamic wormhole
in a slightly different way by a temporal (timelike) outer trapping horizon
\cite{dwh}.
This definition is stricter than the former since it requires 
the horizon to be temporal in order to be locally two-way traversible.
It also gives a unified picture of
wormholes and black or white holes;
the only difference in the definitions 
 is the causal nature of the horizon:
black or white holes are defined by an outer trapping horizon 
which is spatial or null\cite{dwh,bhdyn}.
Thus, dynamic wormholes are defined locally, so that
we need not mention 
the topology of the spacetime as a whole in what follows,
for example two-way traversible means simply that the 
wormhole throat is temporal.
In ref. \cite{dwh}, it is suggested that a stationary
 black hole may become 
a dynamic wormhole by the influx of the negative energy of the Hawking radiation.

As already mentioned above, a wormhole requires negative energy density to maintain
itself, and conversely, one might naively expect that a sufficient condensation
of negative energy makes wormholes. From this point of view we shall ask,
whatever the source of the negative energy,
how much negative energy is needed to make
a wormhole, or what kind of wormholes one can make with less effort.
In the static case, we can conclude that the spherically symmetric
Morris-Thorne wormhole is easiest to make (Sec.~\ref{static}).
In dynamic cases, we can obtain criteria for the existence of
a wormhole in terms of energy densities (Sec.~\ref{dynamic}). 

\section{static wormholes}\label{static}
One may almost uniquely define wormholes 
for static spacetimes in the strict sense, i.e.\ for
spacetimes with a Killing vector field which is everywhere timelike, as 
compact orientable minimal surfaces embedded in the hypersurface 
orthogonal to the Killing vector field.
This definition is a special case of that of dynamic wormholes 
as discussed above.
Since it is convenient to deal with the static case separately,
we briefly discuss this case here, however, results in this section
are not new; for more detailed treatment of static wormholes, see 
ref.~\cite{Hochberg&Visser-2,gwh}. 

We consider here a closed compact oriented minimal surface $H$ embedded in
a 3-manifold $(N,{}^3\g)$.
In what follows, indices run in the ranges $1\le i,j,k,\cdots\le 2$ 
and $1\le A,B,C,\cdots\le 3$.
Let $\{\bei\}$ be an orthonormal basis of $H$, ${\bf e}_3$ its normal, and
$\{\bthi,{\bf \th}^3\}$ the corresponding dual basis.
The structure equations are 
\begin{eqnarray}
&&\d\th^A+\thAB\w\th^B=0\\
&&\frac{1}{2}R^A{}_{BCD}\th^C\w\th^D=\d\thAB+\thAC\w\thCB
\end{eqnarray}
where $\thAB$ is the connection form and $R^A_{BCD}$
the component of the Riemann tensor.
The second fundamental form $\h=h_{ij}\bthi\otimes\bthj$ of $H$ is
given by $h_{ij}=\langle\theta^3{}_i,\bej\rangle$,
 and its trace $\Theta=2^{-1}(h_{11}+h_{22})$
is the mean curvature of $H$.

The area of $H$ is given by $A=\int_H*1$ where $*$ denotes the Hodge operator
with respect to the area form $\th^1\w\th^2$ of $H$.
We slightly deform $H$ by introducing the map $S:H\times I\rightarrow N$,
where $I=[-1/2,1/2]$ is an interval, such that 
 $S$ is an embedding into $N$ for
fixed $t\in I$ denoted by $S_t:H\rightarrow N$,
especially the inclusion map on $H$ for $t=0\in I$, i.e.\ $S_0={\rm id}|_H$.
The metric of $H\times I$ is assumed to be the pullback of ${}^3\g$ by $S$.
Let $\{\bfi,{\bf f}_3\}$ be the orthonormal basis of $H\times I$
 such that $\{\bfi\}$ is tangent to $S_t$ and 
$S_*\bfi|_{t=0}=\bei$. Then its dual basis may be decomposed
as $\omega^i=\th^i+f^i\d t$ and $\omega^3=f\d t$, where $\th^i$
is a one-form on $S_t$, so that now depends on $t$. Correspondingly,
the connection
forms of $H\times I$ are written as $\omega^i{}_j=\theta^i_j$ and
$\omega^3{}_i=\theta^3{}_i+\nabla_{\bfi}f\d t$, where $\nabla$ is the
Riemannian connection of $S_t$. 

The variation of the area
is obtained by differentiating $*1$
with respect to $t$.
We may assume that the deformation is normal to $H$, i.e.\ $f^i=0$.
The first variation of the area form is given by
\begin{equation}
\frac{\p}{\p t}*1|_{t=0}=-*f\Theta,
\end{equation}
so that $\Theta=0$ for $H$ to be a wormhole.

On the other hand, the flare-out condition for 
a static wormhole may be expressed as the stability
of $H$, i.e. the positivity of the second variation
of the area \cite{Chern}. The second variation of the area form
becomes
\begin{eqnarray}
\frac{\p^2}{\p t^2}*1|_{t=0}&=&
*[\nabla_{\bei}f\nabla_{\bei}f-qf^2]
-\d|_H (f\nabla_{\bei}f*\theta^i)\\\nonumber
&=&-*[f\Delta f+qf^2]\\\label{2ndvar}
q&=&\ric(\be3,\be3)+h_{ij}h_{ij}
\end{eqnarray}
where $\ric=R^C{}_{ACB}\theta^A\otimes\theta^B$ denotes the Ricci tensor on
$N$, and $\Delta=*\d|_H*\d|_H$ the Laplacian on $H$. 
We can rewrite $q$ using the Gauss-Codazzi equation as
$2q={}^{(3)}R-{}^{(2)}R+h_{ij}h_{ij}$
where ${}^{(3)}R$ and ${}^{(2)}R$ are the scalar curvatures
 of $N$ and $H$, respectively.
The 2-surface $H$ is stable if and only if the integral of 
eq.(\ref{2ndvar}) over $H$ is non-negative for any function $f$,
or equivalently, the first eigenvalue of the
operator $-\Delta-q$ is positive.
Visser and Hochberg \cite{gwh} defined the strong 
flare-out condition by $q\le 0$ on $H$ which means
$*\p_t^2*1\le 0$ for $f=1$, i.e. the area form of surfaces homogeneously 
deformed from $H$ is not smaller than that of $H$ at each point.
This condition restricts the energy condition according to the
topology of $H$ in a similar way 
to the topology theorem for apparent horizon of time-symmetric
initial data \cite{Gibbons}.
By integrating $q\le 0$ over $H$ using Gauss-Bonnet theorem,
 we obtain the inequality
\begin{equation}
\int_H*(8\pi\mu+\frac{1}{2}\|\h\|^2)\le 2\pi\chi_e
\end{equation}
where $\mu={\bf G}({\bf u},{\bf u})/8\pi$ is the energy density, 
(${\bf G}$ is the Einstein tensor
of the spacetime, ${\bf u}$ the unit vector tangent to the
Killing vector),
 and $\chi_e$ the Euler number of $H$.
In particular, this implies the simple relation
\begin{equation}
U\le \chi_e/4\label{st}
\end{equation}
in terms of the surface energy $U=\int_H*\mu$.
The inequality implies that if $H$ is a double torus or has higher
genus then $\mu$ must be negative somewhere 
on $H$, i.e.\ it implies a violation of the weak energy condition.
If we assume $\mu\ge 0$ by the weak energy condition, 
$H$ is a totally geodesic flat torus
($\mu$ must vanish on $H$ in this case) or a sphere.
Even in these cases, energy conditions are violated,
as is well known from consideration of a different component 
of the Einstein equation, e.g.\cite{Hochberg&Visser-2,gwh}.
The point here is that higher-genus wormholes require 
increasingly large amounts of negative surface energy $U$.
This fact survives in the dynamic case, as discussed in the next section.
Thus, to make a wormhole, it is easier to try for a sphere 
with small principal curvatures,
 e.g.\ a spherically symmetric Morris-Thorne wormhole
\cite{MT}.
\section{dynamic wormholes}\label{dynamic}
Definitions of wormholes for general spacetimes have been given by
Hochberg and Visser \cite{necdwh,dwhats} and by one of the authors 
\cite{dwh}. Though the basic idea is common in both definitions,
i.e.\ the dynamic wormhole is a minimal surface of a null hypersurface,
the latter definition further requires 
the minimal surfaces to generate a temporal hypersurface.
We adopt this quasi-local definition in the following.
Since this definition is slight modification of black or white hole 
definition, some of 
the results below can be obtained by applying known results
for black holes, e.g. Prop.\ref{existence} is an application of ref.~\cite{lightcone}.

The definition is based on the $2+2$ formalism \cite{2+2}, however, we here
adopt the compacted spin-coefficient formalism \cite{PR}.
Consider a foliation of the spacetime $(M,\gd)$ by spatial two-surfaces.
It is convenient to do so by double-null foliation,
i.e.\ a pair of foliations by null hypersurfaces such that
the set of intersections of each null hypersurfaces give a two-dimensional
foliation.
Let $\{\S\}$, $\{\S'\}$ be such one-parameter families of null hypersurfaces, 
and let $S$ be the intersection of $\S\in\{\S\}$ and $\S'\in\{\S'\}$, then
$\{S\}$ is a two-parameter family of spatial two-surfaces.
We here set a spin basis $\{o,\iota\}$ such that $o_A\iota^A=\chi$, 
and a null tetrad $\{\D,\D',\del,\del'\}=\{o\bar o,\iota\bar\iota,
o\bar\iota,\iota\bar o\}$ (normalized
as $\gd(\D,\D')=-\gd(\del,\del')=\chi\bar\chi$, other combinations vanish) 
such that
$\D$ and $\D'$ are
tangent to the future-directed null geodesic generators of
$\S\in\{\S\}$ and $\S'\in\{\S'\}$, respectively, and that
$\{\del,\del'\}$ spans the tangent space of each $S\in\{S\}$.
Its corresponding basis of 
one-forms $\{\gd(\D),\gd(\D'),\gd(\del),\gd(\del')\}$ are denoted by 
$\{\n,\n',\m,\m'\}$, where we may take both $\n$ and $\n'$ to be 
exact one-forms with the freedom of $\chi$.
The spin-coefficients are defined by
\begin{eqnarray}
\nabla_{o\bar o}(o,\iota)&=&(\e o-\k\iota,\gamma'\iota-\t' o)\\
\nabla_{\iota\bar\iota}(o,\iota)&=&(\gamma o-\t\iota,\e'\iota-\k' o)\\
\nabla_{o\bar\iota}(o,\iota)&=&(\b o-\s\iota,\alpha'\iota-\r' o)\\
\nabla_{\iota\bar o}(o,\iota)&=&(\alpha o-\r\iota,\b'\iota-\s' o).
\end{eqnarray}
By construction, $\k=\k'=0$ since $\D$ and $\D'$ are tangent to
geodesics,
$\r-\bar\r=\r'-\bar\r'=0$ since they are orthogonal to  null
hypersurfaces. Moreover, $\e+\bar\e=\t-\bar\alpha-\beta=0$ and
their primed equations hold when $\n$ and $\n'$ are exact,
as occurs if the spin-basis is adapted to the null hypersurface such that 
$\n$ and $\n'$ are the differentials of the null coordinates;
this choice of gauge is discussed in ref.\cite{sc}.

A trapping horizon is defined to be 
a one-parameter family of two-surfaces $\{H\}$
on which one of the contracting rates, say $\r$, vanishes.
Each $H$ locally determines 
two (ingoing and outgoing) normal null hypersurfaces, 
so that we adopt the natural double-null foliation generated by $\{H\}$,
i.e.\ such that $\{\del,\del'\}$ spans $H$.
The trapping horizon is
 called {\em future} if $\r'>0$, and {\em past} if $\r'<0$.
Moreover, trapping horizons are classified according to the
contracting rates of neighbouring light-cones: they are called
{\em outer} if $\thorn'\r>0$ and {\em inner} if $\thorn'\r<0$.
Then a black (white) hole is defined to be a outer-future (-past)
 trapping horizon which is null or spatial.
Note that the notion of a black hole is invariant once a double-null
foliation has been set up, since the product $\r\r'/\chi\bar\chi$ or
$\thorn'\r/\chi\bar\chi$ is invariant under the
transformation of the basis: $\{o,\iota\}\mapsto\{\lambda o,\mu\iota\}$
by complex functions $\lambda$ and $\mu$, which preserves the
double-null foliation.
The condition that the trapping horizon is null or spatial in the definition of
a black or a white hole is guaranteed if the null energy condition holds, i.e. 
the stress-energy tensor ${\bf T}$ satisfies
${\bf T}({\bf k},{\bf k})\ge 0$ for any null vector ${\bf k}$.
To see this, introduce a vector ${\bf z}=a\D+b\D'$ which belongs to the 
orthogonal complement of $\{\del,\del'\}$ in the tangent space of the
outer trapping horizon. The condition ``outer'' requires
$\D'\r>0$ on the horizon, while the spin-coefficient equation \cite{PR}
\begin{equation}
\thorn\r=\r^2+\s\bar\s+\Ph00\label{focus}
\end{equation}
implies $\D\r\ge 0$ since $\Ph00\ge 0$ by the null energy condition, so
that $ab\le 0$ since ${\bf z}$ is tangent to the horizon: 
${\bf z}\r=a\D\r+b\D'\r=0$, as required, which implies 
$\gd({\bf z},{\bf z})=2ab\le 0$, i.e.\ the outer trapping horizon
is spatial or null.

Similarly, a wormhole horizon 
is defined to be an outer trapping horizon which
is temporal \cite{dwh}, 
however,
it should be noted that we have excluded
the degenerate case: $\thorn'\r=0$ for simplicity,
in which case additional technical
complications must be discussed \cite{dwhats}.
To see the validity of the definition, we shall prove the following
lemma.

\begin{lem}
Let $\{H\}$ be a trapping horizon, $\r=0$, then any two of the
following conditions imply the third:\\
(a) $\{H\}$ is temporal;\\
(b) $\{H\}$ is outer, $\thorn'\r>0$;\\
(c) $\thorn \r<0$.
\label{temporal}
\end{lem}
\noindent
{\em Proof.} 
Let ${\bf z}=a\D+b\D'\ne 0$ be a tangent vector of $\{H\}$.
Assume the condition (a) holds, which means $ab>0$, then the
equivalence of (b) and (c) comes from 
${\bf z}\r=a\thorn\r+b\thorn'\r=0$. Assume instead (b) and (c), then
we obtain $ab>0$ which implies (a).\hfill $\Box$
\medskip

\noindent
Thus, a wormhole horizon is composed of minimal surfaces of null hypersurfaces
in the sense $\thorn\r<0$.
The Hochberg-Visser wormhole definition requires (c) but not (a) or (b).
By virtue of the simplicity of the definition, 
one may immediately obtain the negative energy theorem:
$\Ph00<0$, i.e.\ the null energy condition is violated on the wormhole
horizon.
The energy-momentum of any reasonable classical matter, even of 
the cosmological constant, satisfies the null energy condition, so that
it is common to appeal to quantum effects such as the Casimir effect or
the Hawking radiation for the negative energy source of wormholes.

In the following, we consider how much negative energy is needed to
construct a wormhole, whatever the negative energy is.
We may precisely answer this question in terms of the two components of the Einstein equation which contain $\thorn\r$ and $\thorn'\r$.
Firstly we integrate the focusing
equation on the null hypersurface.
In ref.\cite{lightcone}, the focusing equation is integrated on a
light-cone, and we take a similar
procedure with a different boundary condition.
Consider a null hypersurface $\S\in\{\S\}$, and let $\D$ be tangent
to the null geodesic generators of $\S$.
Take a section $S_0$ of $\S$ which is assumed to be
a compact two-surface without boundary,
 and on which the light ray is assumed to
be contracting in the $\D$-direction, i.e.\ $\r>0$ on $S_0$.
One could also set such a boundary condition at past null infinity 
in an asymptotically flat space-time; 
the condition is just meant to imply that $\D$ is an ingoing direction.
The spin-coefficient equation (\ref{focus}) may be written  
more explicitly as
\begin{equation}
\D\r=(\e+\bar\e)\r+\r^2+\s\bar\s+\Ph00.\label{focus2}
\end{equation}
When $\D$ is tangent to an affinely parametrized geodesic,
$\e$ is pure imaginary, so that the first term of the r.h.s.\
vanishes, which might be the most usual choice of $\D$, however,
we here take a different choice of $\D$.
The boost transformation
\begin{equation}
o\mapsto ao
\end{equation}
with a real number $a$,
preserves the null hypersurface $\S$.
By this transformation, the spin-coefficients
$\r$ and $\e$ transform to $a^2\r$ and $a^2(\e+\D\ln a)$,
respectively.
Now let $a$  be a solution of the differential equation,
\begin{equation}
2\D\ln a+\r+\e+\bar\e=0,
\end{equation}
then the sum $\r+\e+\bar\e$ vanishes by the boost with $a$, so that 
the eq.(\ref{focus2}) takes a simple form
\begin{equation}
\frac{\p}{\p u}\r=\s\bar\s+\Ph00
\end{equation}
where $u$ is the new parameter of the null generator 
of $\S$ such that $\p/\p u$ coincides with $\D$, and
$u=0$ on $S_0$ (note that caustics in the future of $S_0$, if they
exist, are pushed into infinite $u$ in this parametrization).
Then, $\r$ is written in the form
\begin{equation}
\r=\r_0+\int_0^{u}(\s\bar\s+\Ph00)\d u\label{rholambda}
\end{equation}
where $\r_0>0$ is the contracting rate on $S_0$ in this gauge.
Eq.(\ref{rholambda}) implies the existence of a minimal surface
by the sufficient influx of the negative energy in the $\D$-direction
(note that the existence of a single minimal surface implies that
 there will be locally a trapping horizon consisting 
of such surfaces by continuity
of the function $\r$).
Especially, we obtain the existence theorem of the trapping horizon;

\begin{prop}
If $\Ph00\le-(|\s|^2+\varepsilon)$, $(0<u<\r_0/\varepsilon)$
holds for some positive constant $\varepsilon$,
then there exists a trapping horizon.
\label{existence}
\end{prop}

\noindent
However, this proposition does not say whether or not the trapping horizon
is a wormhole horizon as defined here; 
we need to satisfy (a) or (b) as well as (c) above.
To state this in an invariant way, we transform the basis as follows.
If the assumption of the Prop.\ref{existence} is satisfied,
then this trapping horizon in general will not be spaned by 
$\{\del,\bar\del\}$, but we are able to 
make $\{\del,\bar\del\}$ span the trapping
horizon by the transformation:
\begin{eqnarray}
\iota\mapsto\iota+Eo
\end{eqnarray}
with an appropriate complex function $E$.
Note that all the quantities used in the Prop.\ref{existence} are
unchanged by this transformation, though the new tetrad need not be
associated with the original double-null foliation.
In this setting, we state the following lemma.

\begin{lem}
Let $H$ be the trapping horizon in the Prop.\ref{existence}, and
let $k$ be the Gaussian curvature of $H$, then $H$ is a wormhole
if and only if the inequality
\begin{equation}
(\Ph11+3\Pi)< 2^{-1}\chi\bar\chi k-\t\bar\t-\Re[\eth'\t]\label{dominant}
\end{equation}
holds on $H$.
\end{lem}
\noindent
{\em Proof.}
It is easily seen that $\thorn\r<0$ holds on $H$
in the situation of the Prop.\ref{existence}.
On the other hand, 
the real part of the following spin-coefficient equation \cite{PR}
\begin{equation}
\thorn'\r-\eth\t=\r\bar\r'+\s\s'-\t\bar\t-\Psi_2-2\Pi
\end{equation}
can be rewritten as
\begin{equation}
\thorn'\r=2^{-1}\chi\bar\chi k-\t\bar\t-\Re[\eth'\t]-(\Ph11+3\Pi)
\end{equation}
on $H$, since $\r=0$ on $H$, and the sectional curvature $k$
of the distribution $\{\del,\del'\}$ can be in general written as
$k=2\Re[\s\s'-\r\r'-\Psi_2+\Pi+\Ph11]/\chi\bar\chi$.
Hence eq.(\ref{dominant}) is equivalent to $\thorn'\r>0$ on $H$,
and moreover, $\{H\}$ is temporal by the lemma~\ref{temporal}.\hfill $\Box$
\medskip

Conversely, the inequality (\ref{dominant}) must be satisfied on
any wormhole horizon.
Especially, by integrating the inequality (\ref{dominant}) divided by
$\chi\bar\chi$ over
the two-surface $H$ spaned by $\{\del,\del'\}$ with the area element
$*1=(i/\chi\bar\chi)\m\w\m'$, we obtain an additional constraint
\begin{equation}
W<\chi_e/4\label{intdominant}
\end{equation}
in terms of the surface energy
\begin{equation}
W={1\over{4\pi}}\int_H*\frac{\Ph11+3\Pi}{\chi\bar\chi}\label{surface-energy}
\end{equation}
and the Euler number $\chi_e$ of $H$;
$\chi_e=2(1-g)$ where $g$ is the genus (number of handles).
This follows from the Gauss-Bonnet theorem
\begin{equation}
\int_H*k=2\pi\chi_e
\end{equation}
and the fact that $\Re[\eth'\t]/\chi\bar\chi$
is a total divergence.
Thus, the energy density $(\Ph11+3\Pi)/4\pi$ must be negative somewhere on $H$,
and the total surface energy $W$ must be negative,
unless $H$ has spherical topology.
This is a dynamic version of the static inequality (\ref{st}).
Note that equality has been excluded here 
by assuming the strict inequalities (b) and (c) in the definition of wormhole,
i.e.\ excluding degenerate wormholes.
Degenerate cases may be included, as for the topology law 
of black or white holes \cite{bhdyn,sc}, 
which involves the same method as above.
 
Inequality (\ref{intdominant}) can be sharpened 
in terms of the surface gravity \cite{1st,ql1}
\begin{equation}
\kg={1\over{4\pi r}}\int_H*{\thorn'\r\over{\chi\bar\chi}}
\end{equation}
where $r=\sqrt{A/4\pi}$ in terms of the area $A=\int_H*1$ of $H$.
Then
\begin{equation}
W\le\chi_e/4-r\kg.\label{sharpened}
\end{equation}

Finally, we prove the existence theorem of a wormhole with 
spherical symmetry to obtain a more explicit estimate of the negative energy,
which also provides a lower bound for the horizon area.

\begin{prop}
Let $(M,\gd)$ be spherically symmetric, and
let the double-null foliation by $\{\S\}$ and $\{\S'\}$
respect the spherical symmetry.
Let $\D$ be the tangent vector of the affinely parametrized null
geodesic generator of $\{\S\}$, 
and let $S_0$ be a spherically symmetric 2-sphere on a light-cone
$\S\in\{\S\}$ with area $A_0$ and 
contracting rate $\r_0>0$.

If $\Phi_{00}<-(p\r_0)^2$ holds for some constant $p>1$ on
$\S$ during the affine distance 
$r_1=(2p\r_0)^{-1}\ln[(p+1)/(p-1)]$ from  $S_0$ in the $\D$-direction,
then there exists a trapping horizon $H$ on $\S$ with area 
$A>\sqrt{1-p^{-2}}A_0$, and moreover, $H$ is a wormhole horizon
if and only if $A^{-1}>(\Ph11+3\Pi)/2\pi\chi\bar\chi$ holds.
\label{spherical}
\end{prop}

\noindent
{\em Proof.}
By construction, $\D$ is tangent to affinely parametrized geodesic
$\k=\e+\bar\e=0$,
twist-free $\r=\bar\r$, sheer-free $\s=0$, and non-rotating $\t=0$.
Then, eq.(\ref{focus}) becomes
\begin{equation}
\frac{\p}{\p r}\r=\r^2+\Phi_{00}.
\end{equation}
We have by integration
\begin{equation}
\r<p\r_0\tanh [p\r_0 (r_1-r)]
\label{expansion}
\end{equation}
for $0\le r\le r_1$ ($r=0$ on $S_0$).
Since the r.h.s.\ of eq.(\ref{expansion}) vanishes
for $r=r_1$, which implies $\r|_{r=r_1}<0$, there exists a
trapping horizon $\r=0$ for $r=r_2$, $(0<r_2<r_1)$.
The lower bound of the horizon area is obtained by
integrating the equation $\p A/\p r=-2\r A$, and the condition that $H$
is a wormhole horizon is obtained from the inequality
(\ref{dominant}).\hfill $\Box$
\medskip

\section{Discussion}
We have shown that locally defined wormholes generally require constraints 
on the energy density 
in two independent components of the stress-energy tensor,
corresponding to two independent components of the Einstein equation.
One of these is familiar from previous work \cite{necdwh}--\cite{dwh}, 
as a necessary condition 
for the existence of a minimal surface in a null hypersurface.
We have also integrated the relevant equation along the null hypersurface 
to obtain a sufficient condition for a minimal surface to form,
in terms of sufficient negative energy density crossing the hypersurface.
For this to be part of a locally traversible wormhole horizon requires 
the second condition,
the inequality (\ref{dominant}).
This can be satisfied without the relevant energy density becoming negative 
for minimal surfaces of spherical topology,
but higher-genus wormholes require the total surface energy $W$ 
to be strictly negative.
In this sense, higher-genus wormholes are less likely to form.

Note that these conditions constrain the negative energy density 
or its surface integral over the wormhole, but not its volume integral, 
i.e.\ the total negative energy in a region.
Theoretically one could arrange for the negative energy density to be confined 
to a small region around the wormhole, 
surrounded by large quantities of positive energy 
yielding a positive total energy which is as large as one pleases.
Even if one integrates only over the region where the energy violations occur,
this region may be arbitrarily small, 
leading to an arbitrarily small total negative energy.
Such thin shells can even satisfy the quantum inequalities mentioned below.

As an example, 
for static, spherically symmetric (i.e.\ Morris-Thorne) wormholes,
the tension minus energy density is just $\kg/2\pi r$.
Thus the required negative energy density is smaller 
for large wormholes with small surface gravity,
both of which are practical requirements for a comfortably traversible wormhole.
The corresponding negative energy, 
integrated over a shell of width $\ell$ around the wormhole, 
is of the order of $2\ell r\kg$, so increases with wormhole size, 
but is still smaller for small surface gravity.

Moreover, the local energy densities do not directly determine 
the sign of active gravitational energy, 
such as the asymptotic (ADM and Bondi) energies 
in an asymptotically flat space-time, 
or quasi-local energies such as the Hawking energy
or (in spherical symmetry) the Misner-Sharp energy \cite{1st,ss}.
The Hawking energy is actually positive 
on a wormhole horizon with spherical topology, 
taking the value $\sqrt{A/16\pi}$.
The asymptotic energies may also be positive in a wormhole spacetime.
Rather, the sign of the energy density affects 
the derivative of the active energy \cite{ss,mono,charge}.
For instance, 
the usual Bondi energy-loss property reverses 
under negative energy---specifically, 
the null energy condition with reversed sign---leading to an increase 
of Bondi energy at future null infinity.
These perhaps counter-intuitive properties of negative energy density 
should be considered in regard to beliefs such as that 
wormholes are likely to be plagued by naked singularities,
which are associated with negative (active) energy, e.g.\cite{ss}.
The increase of the Bondi energy actually suggests the opposite,
that there might be a form of cosmic censorship for wormholes:
wormholes are either stable or collapse to black holes.
Of course, this will depend on the matter model e.g.\cite{dwh}
and perhaps also on the initial configuration.

The results suggest that an advanced civilization would be able to construct a
traversible wormhole if it could prepare the required negative energy.
On the other hand, though quantum field theory permits negative energies,
it also constrains the magnitude and duration of the negative energy,
according to so-called quantum inequalities, 
uncertainty principle-type relations.
Applying quantum inequalities to static spacetimes,
Ford and Roman \cite{FR} discussed that there will be at
best wormholes of Planckian size,
though in fact this allows large, thin-shell wormholes.
Also, there seems to be more room for debating
 dynamic wormholes.
Prop. \ref{spherical} implies that 
a Schwarzschild black hole
requires an arbitrarily small amount of negative energy near its horizon
to become a dynamic wormhole,
so that it seems that such a wormhole can exist due to the
Hawking radiation.
Since the temperature of the Hawking radiation of a black
hole of macroscopic size is low,
the horizons will be moving apart at close to the speed of light,
so that the resulting wormhole would be difficult to traverse in practice,
though two-way traversible in principle. 
However, the possibility of the existence
of such  macroscopic wormholes itself has theoretical
importance.

Throughout, we have discussed the local structure of wormholes,
without prejudice as to whether there will be another 
universe or a distant region of the spacetime beyond the
wormhole throat, or whether the hypothetical advanced civilization
could identify spacetime regions as relativists or
topologists do.

{\em Acknowledgements.}
We would like to acknowledge many helpful discussions with
Prof.\ Humitaka Sato, Dr.\ Ken-ichi Nakao,
Dr.\ Masaru Siino, and Dr.\ Shigeki Sugimoto.
Helpful comments by Prof. Matt Visser are gratefully acknowledged.

\end{document}